# Atomic Scale Memory at a Silicon Surface

R. Bennewitz[*], J. N. Crain, A. Kirakosian, J.-L. Lin, J. L. McChesney, D. Y. Petrovykh, and F. J. Himpsel

Dept. of Physics, UW-Madison, 1150 University Ave., Madison, WI 53706, USA

[*] Dept. of Physics and Astronomy, University of Basel, 4056 Basel, Switzerland

## Abstract

The limits of pushing storage density to the atomic scale are explored with a memory that stores a bit by the presence or absence of one silicon atom. These atoms are positioned at lattice sites along self-assembled tracks with a pitch of 5 atom rows. The memory can be initialized and reformatted by controlled deposition of silicon. The writing process involves transfer of Si atoms to the tip of a scanning tunneling microscope. The constraints on speed and reliability are compared with data storage in magnetic hard disks and DNA.

In his landmark 1959 talk at Caltech, Richard Feynman estimates that "all of the information that man has carefully accumulated in all the books in the world, can be written … in a cube of material one two-hundredth of an inch wide". Thereby, he uses a cube of 5x5x5=125 atoms to store one bit, which is comparable to the 32 atoms that store one bit in DNA. Such a simple, back-of-the-envelope calculation gave a first glimpse into how much room there is for improving the density of stored data when going down to the atomic level. In the meantime, there has been great progress towards miniaturizing electronic devices all the way down to single molecules or nanotubes as active elements (*1*). Memory structures have been devised that consist of crossed arrays of nanowires linked by switchable organic molecules (*2,3*) or crossed arrays of carbon nanotubes with electrostatically switchable intersections (*4*).

It is our goal to push the storage density to the atomic limit and to test whether a single atom can be used to store a bit at room temperature. How closely can the bits be packed without interacting? What are the drawbacks of pushing the density to its limit while neglecting speed, reliability, and ease of use? The result is a two-dimensional realization of the device envisaged by Feynman, as shown in Fig. 1. A bit is encoded by the presence or absence of a Si atom inside a unit cell of 5x4=20 atoms. The remaining 19 atoms are required to prevent adjacent bits from interacting with each other, which is verified by measuring the autocorrelation. A specialty of the structure in Fig. 1 is the array of self-assembled tracks with a pitch of 5 atom rows that supports the extra atoms. Such regular tracks are reminiscent of a conventional CD-ROM (*5*). However, the scale is shrunk from µm to nm, which corresponds to a million times higher density. The readout of such a memory via scanning tunneling microscopy (STM) is obvious, albeit slow. Writing is more difficult. While atoms can be moved controllably at liquid helium temperature (*6*), it is much harder to achieve that at room temperature (*7-14*). In order to prevent diffusion it is necessary to choose atoms that are strongly bound to the surface. Moving them requires strong forces and a close approach by the STM tip, which entails the risk of an atom jumping over to the tip. We use this effect as virtue to remove a silicon atom from the surface for writing a 0. The memory is pre-formatted with a 1 everywhere by controlled deposition of silicon onto vacant sites.

In the following we begin with a description of the memory structure, move on to writing and reading, and eventually explore reliability and speed. The outlook considers the fundamental limitations of a single atom memory and makes a comparison to data storage in DNA.

The self-assembled memory structure shown in Figs. 1,2 is obtained by depositing 0.4 monolayers of gold onto a Si(111) surface at $700^0$C with a post-anneal at $850^0$C, thereby forming the well-known Si(111)5x2−Au structure (*15-17*). All images are taken by STM with a tunneling current of 0.2nA and a sample bias of −2V. At this bias the extra silicon atoms are enhanced compared to the underlying 5x2 lattice. A stepped Si(111) substrate tilted by $1^0$ towards the [ $\bar{1}$ $\bar{1}$ 2 ] azimuth is used to obtain one of the three possible domain orientations exclusively (*17*). The surface arranges itself into tracks that are exactly 5 atom rows wide (Fig. 1 right). They are oriented parallel to the steps in the [ $\bar{1}$ 1 0 ] direction. Protrusions reside on top of the tracks on a 5x4 lattice. Only half of the possible sites are occupied in thermal equilibrium (Fig. 2a). When varying the Au coverage the occupancy remains close to 50%. Excess Au is taken up by patches of the Au-rich Si(111)√3x√3-Au phase, and Au deficiency leads to patches of clean Si(111)7x7. In order to find out whether the protrusions are Si or Au we evaporate additional Si and Au at low temperature ($300^0$C). Silicon fills the vacant sites (Fig. 2b,d), but gold does not. In Fig. 2b the occupancy of the 5x4 sites has increased to 90%±3% from 53%±4% in Fig. 2a. Higher annealing allows the extra Si to diffuse away to the nearest step and causes vacancies to reappear, confirming that the half-filled structure is thermodynamically stable. Thus, an average code with 1 and 0 in equal proportion is particularly stable.

The writing process consists of removing Si atoms from a nearly-filled lattice, such as that in Fig. 2b,d. Figure 3 demonstrates one of two methods, which is based on chemical attachment to the tip. The tip is brought down towards the Si atom to be removed, typically by 0.6nm for 30ms without applying a voltage. This method is similar to previous work on Ge(111), where 98% reliability has been achieved in removing a single atom (*13*). A less reliable method uses field desorption by a voltage pulse of −4V on the sample (30ms long) with the tip hovering above the Si atom to be removed.

The readout is demonstrated in Figure 2e. A line scan along one of the tracks in Fig. 2c (marked by an arrow) produces well-defined peaks for extra Si atoms that protrude well beyond the noise level. Since the memory is self-formatted into tracks it can be read by a simple, one-dimensional scan. There is no need to search in two dimensions for the location of a bit. The signal is highly predictable since all atoms have the same shape and occur on well-defined lattice sites. After subtracting identical Gaussians at the lattice sites one obtains a residual comparable to the noise (Fig. 2e bottom trace). The height of the signal (z=0.13nm) exceeds the noise ($\delta$z=0.005nm rms) by a factor of 26, using a dwell time of 500µs per point. A highly reproducible pulse shape allows sophisticated signal filtering techniques because most noise signals do not match the known shape and can be removed. An example is partial response maximum likelihood detection (PRML), which is widely used for the readout of magnetic hard disks (*18-20*) and in long-distance communications. In PRML the signal is filtered to produce a standard line shape, sampled at regular intervals, and processed in real time by the Viterbi algorithm, which selects the most likely bit sequence.

Reliability becomes a key issue with such a small memory cell. The writing process is too slow and error prone to be practical, but the readout deserves closer inspection. Our data provide quantitative input for determining the achievable error rate, the thermal stability, and correlations between adjacent bits.

The error rate for reading is related to the effective signal-to-noise ratio (SNR), which is determined by three input parameters for hard disk readout using PRML (*18*):

$$\mathrm{SNR} = 2/\pi \cdot \mathrm{WB}/\sigma^2$$

W is the full width half maximum of a signal pulse, B the bit spacing, and $\sigma$ the variance in the jitter of the pulse positions. Typical values for hard drives (*19*) are W≈120nm, B≈50nm, $\sigma$≈4nm, giving SNR≈240≈24dB and an error rate of $10^{-8}$. For the atomic memory one can derive analogous quantities W=0.55nm, B=1.54nm, and $\sigma$=0.015nm from Fig. 2e by taking the peak width, the lattice spacing, and the jitter of the peak positions relative to the lattice points (from Gaussian fits with unconstrained positions). These numbers can serve as input for designing filters and codes that minimize the error rate. Such models will be different from those for hard disks, where readout pulses alternate in sign and bits are less than a pulse width apart. A closer analog might be the unipolar soliton pulses that are used in long distance communications through optical fibers. Pulse amplitude variation plays a role for solitons, which is 0.005nm or 4% of the peak height in our case.

The thermal stability of Si atoms on the tracks can be estimated by fitting high temperature STM results from (*15*) to a simple model of activated diffusion. The temperature dependence of the jump rate ν(T) is determined by a Boltzmann factor containing an activation energy ΔE:

$$\nu(T) = \nu_0 \exp(-\Delta E/k_B T)$$

The attempt frequency $\nu_0$ is comparable to the frequency of lattice vibrations, with a typical value $\nu_0=10^{13} s^{-1}$. Estimating a jump rate of about $1 s^{-1}$ at a temperature T=475K from the data in (*15*) one obtains an activation energy $\Delta E = k_B T \ln(\nu_0/\nu) \approx 1.2 eV$. That gives a jump rate of $10^{-8} s^{-1}$ at room temperature ($k_B T=25 meV$), i.e., one jump in 2-3 years. Thus, thermal stability is not an issue compared to less fundamental limits, such as surface contamination.

Correlations between adjacent bits come into play at high storage density. In magnetic storage one has to be concerned about magnetic coupling between adjacent particles at a bit spacing of 10nm or less (*20-22*). The bits are much closer than that on the Si surface (1.5nm along the track and 1.7nm between tracks). In order to detect interactions between the Si atoms we have determined the autocorrelation function of their equilibrium distribution (Fig. 4). Along the tracks, the nearest site of the underlying 5x2 lattice is almost completely excluded, with an occupancy of only 0.04 relative to the average (Fig. 4 bottom, compare (*16*)). Therefore, a bit spacing of 2 lattice sites would discriminate a 1 1 pair against a 1 0 by a factor of 25. All other correlations are much closer to the average of 1, with the largest deviation (1.33) occurring at the second 5x2 site in Fig. 4 (bottom). Thus, the 5x4 cell represents the smallest viable cell for the underlying 5x2 lattice that keeps bit interactions under control. Feynman's proposed spacing of 5 atoms between bits was right on the mark.

One of the fundamental limitations to devices operating on the atomic scale is speed (Figure 5). For example, the minimum switching time t is given by the uncertainty relation t=h/E, where E is the switching energy (*23*). E has to be larger than the minimum energy $E_{min}= k_B T \cdot \ln 2$ for switching one bit (*23,24*). In our case, the activation energy of 1.2eV for moving one Si atom is much larger than $k_BT=25$meV. In principle, that would allow very fast switching and writing.

The readout, however, has to slow down for small bits since the signal decreases and becomes noisier. In our case, the tunneling current is affected by statistical fluctuations in the number of electrons and by thermal noise. Their respective spectral densities are $S_s(\omega)=2eI$ and $S_t(\omega)=4k_BT/R$, resulting in current fluctuations of 8fA/√Hz and 1.3fA/√Hz for our conditions (I=0.2nA, R=$10^{10}\Omega$, T=300K). Adding the two noise contributions and integrating over a dwell time of $\tau=500\mu$s per point one finds a current fluctuation $\delta I = [(S_s+S_t)/\tau]^{1/2} = 3.6\cdot10^{-13}$A. This current fluctuation is translated into a height fluctuation $\delta z$ via the exponential dependence $I(z) = I_0 \exp(-kz)$ of the tunneling current on z:

$\delta z = \delta I/(\partial I/\partial z) = \delta I/(-kI) = 9\cdot10^{-5}$nm

with k=20nm$^{-1}$ for a typical tunnel barrier of 4eV. That is 55 times smaller than the actual noise $\delta z$=0.005nm. Statistical and thermal noise would meet this level with a dwell time of only 160ns (200 electrons per point). The corresponding readout speed would be $6\cdot10^6$ points/s, which is respectable but still slower than today's hard disks (Fig. 5). High speed STM amplifiers operating at rates up to 50MHz exist (*25*).

A further tool for enhancing speed is a high degree of parallelism (see the discussion in (*24*) and (*26*)). In fact, there has been a substantial effort directed at producing large arrays of scanning probe tips by silicon processing methods. An array of 32x32=1024 tips with 92$\mu$m pitch is operational (*27*). The atomic precision of the tracks ensures that the tip array follows the tracks after a one-time adjustment of the tip positions and the scan direction.

An interesting yardstick is the storage and transcription of data in biological systems. 5x4=20 surface atoms store one bit on silicon compared to 32 atoms used by DNA (64 atoms for a AT base pair plus back-bone, 63 atoms for CG, with each base pair coding the four combinations AT, TA, CG, GC, i.e. 2 bits). The transcription rate from DNA to RNA is ≈60nucleotides/s for E-coli at 37$^0$C, and 10 times faster for DNA replication. The STM acquisition rate on silicon is comparable (120bit/s in Fig. 2e). It could be as high as $10^7$bit/s at the statistical noise limit. Parallel readout can be used in both cases, with ≈$10^1$ sub-sections of DNA being replicated simultaneously and an array of $10^3$ tips scanning in parallel. Cells use a similar parallelism of $10^3$–$10^4$ for protein synthesis, where speed is more important. The error rate achieved in DNA replication is as low as $10^{-7}$–$10^{-11}$ with error correction by DNA polymerase.

Compared to conventional storage media both DNA and the silicon surface excel by their storage density (compare Fig. 1 for CD-ROM). The highest density achieved in hard disk demos is about 100Gbit/inch$^2$, whereas the Si atom memory exhibits 250Tbit/inch$^2$. However, the push towards the atomic density limit requires a sacrifice in speed, as demonstrated in Fig. 5. Practical data storage might evolve in a similar direction, with the gain in speed slows down as the density increases. Somewhere on the way to the atomic-scale ought to be an optimum combination of density and speed.

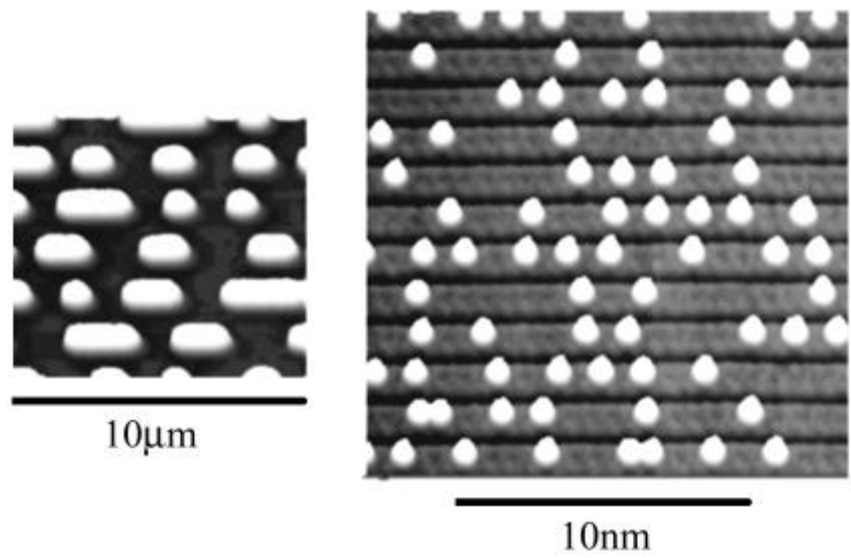

**Fig. 1.** Comparison of the atomic memory on silicon with a CD-ROM [5]. Extra silicon atoms occupy lattice sites on top of tracks that are 5 atom rows wide (1.7nm). The scale is reduced from μm to nm, which leads to a $10^6$ times higher density.

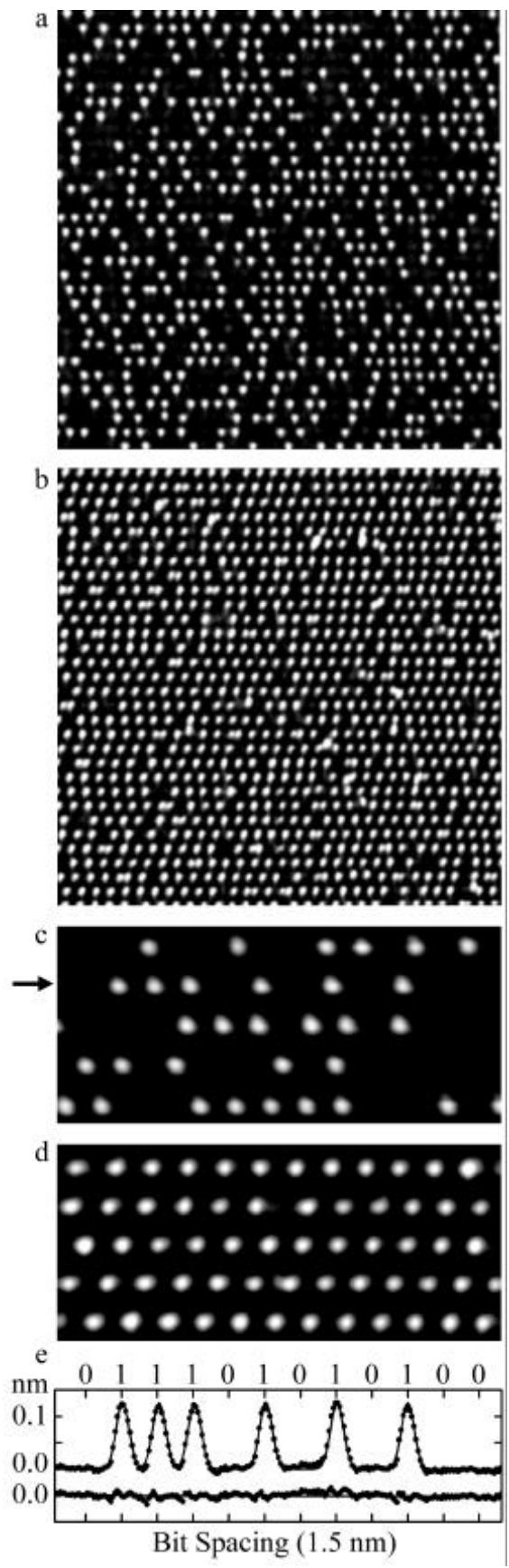

**Fig. 2.** STM images of the equilibrium structure with half filling of the atomic sites (a,c) and the pre-formatted structure with nearly complete filling (b,d). A line scan with 500μs dwell time per point is shown in (e), following the arrow in (c). Identical peaks have been subtracted at the lattice sites of the Si atoms in the bottom curve to demonstrate a highly-reproducible signal.

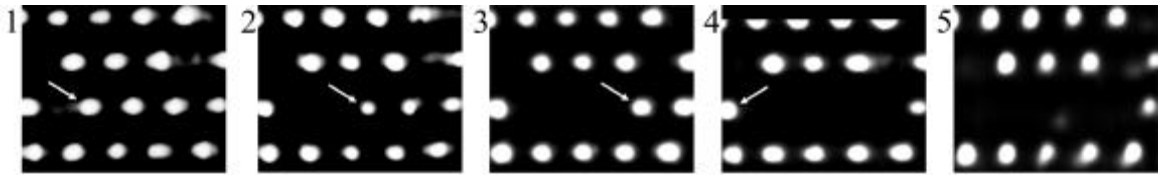

**Fig. 3.** Writing a sequence of four zeros in one row (labeled by arrows). Silicon atoms are transferred to the STM tip.

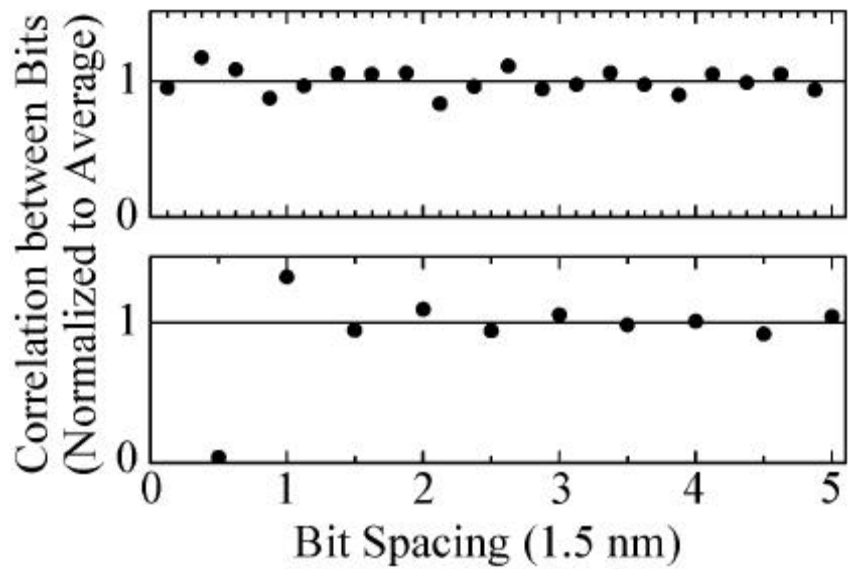

**Fig. 4.** The autocorrelation between neighboring Si atoms as a test of interactions between adjacent bits. The closest 5x2 site along a track is nearly excluded (bottom), showing that 5x4 atoms is the minimum viable cell size for this lattice. Neighboring tracks are not correlated (top).

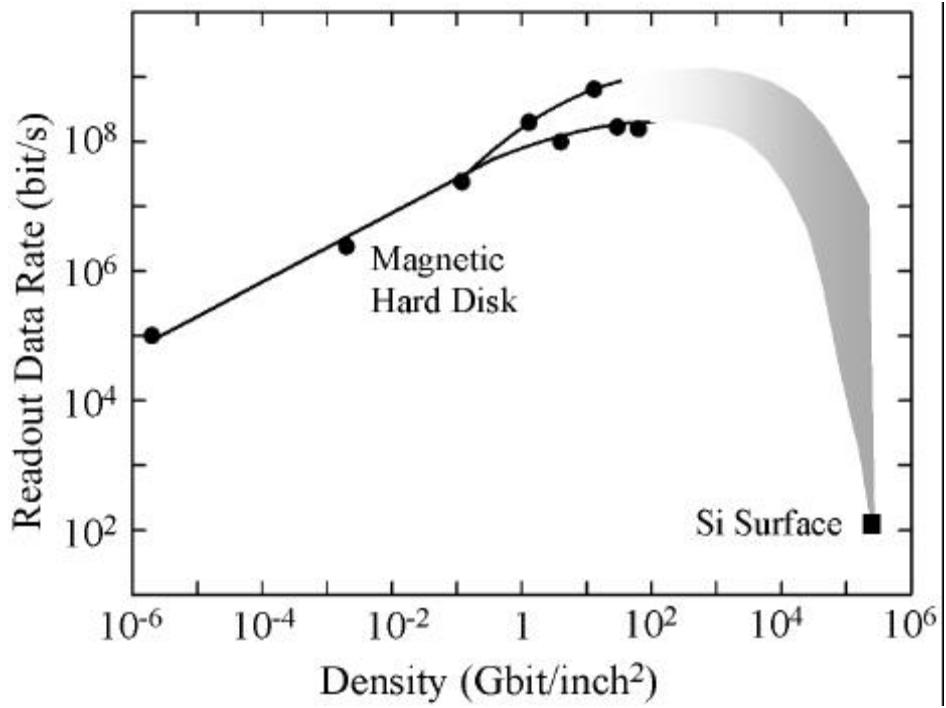

**Fig. 5.** Tradeoff between readout speed and storage density approaching the atomic limit.